\documentclass{iaus_selma}
\usepackage{graphicx}
\usepackage{natbib}
\newcommand{\Msun}{\mbox{$~\mathrm{M}_{\odot}$}}

\def\aap{\mbox{\bf{Astron.\ Astrophys.}}}
\def\apj{\mbox{\bf{Astrophys.\ J.}}}
\def\apjl{\mbox{\bf{Astrophys.\ J.\ L.}}}
\def\aj{\mbox{\bf{Astron.\ J.}}}

\def\pasp{\mbox{\bf{Pub.\ Astron.\ Soc.\ Pacific}}}
\def\mnras{\mbox{\bf{Mon.\ Not.\ Roy.\ Astron.\ Soc.}}}

\def\araa{\mbox{\bf{Ann.\ Rev.\ Astron.\ Astrophys}}}

\title[]
{Massive binaries and the enrichment of the interstellar medium in globular clusters}

\author[]
{S.E. de Mink$^1$ \and  O.R. Pols$^1$ \and  N. Langer$^{2,1}$ \and  R.G. Izzard$^3$}

\affiliation{
$^1$Astronomical Institute Utrecht, Postbus 80000, 3508 TA, Utrecht, the Netherlands\\
$^2$Argelander-Institut   f\"{u}r Astronomie, Bonn, Auf dem H\"{u}gel 71,
   53121 Bonn, Germany \\
$^3$ Universit\'e Libre de Bruxelles, Boulevard du Triomphe, B-1050 Brussels, Belgium \\
 email: {\tt S.E.deMink@uu.nl} 
}

\pubyear{2009}
\volume{266}  
\pagerange{x-x}
\jname{Star Clusters: Basic Galactic Building Blocks Throughout Time And Space}
\editors{eds.}
\begin{document}
\maketitle

\begin{abstract}
Abundance anomalies observed in globular cluster stars
 indicate pollution with material processed by hydrogen
 burning.  Two main sources have been suggested: asymptotic giant
     branch (AGB) stars and massive stars rotating near the break-up
     limit (spin stars). We discuss the idea put forward by \citet{DeMink+09b} that massive binaries may provide an interesting alternative source of processed material.

 We discuss observational evidence for mass shedding from interacting binaries.  In contrast to the fast, radiatively driven winds of massive stars, this material is typically ejected with low
     velocity. We expect that it remains inside the potential well of
     a globular cluster and becomes available for the formation or
     pollution of a second generation of stars.
     We estimate that the amount of processed low-velocity material
     that can be ejected by massive binaries is larger than the contribution of two previously suggested sources combined.

\end{abstract}

\firstsection 
\section{Introduction}

For a long time star clusters have been considered as idealized single-age, chemically homogeneous stellar populations. However, it has recently become clear that some clusters show multiple main sequences and sub-giant branches and extended horizontal branches \citep[e.g.][]{Piotto+07}.  The presence of these features in some intermediate-age clusters may be explained by the effects of stellar rotation on the location of stars in the colour magnitude diagram \citep{Bastian+DeMink09}. However, this provides no explanation for the features seen the old globular clusters, which seems to imply the existence of multiple populations within one cluster.

In addition, large star-to-star abundance variations are found for
light elements such as C, N, O, Na and Al, while the composition of
heavier elements (Fe-group and $\alpha$-elements) seems to be
constant.  Field stars with the same metallicity do not exhibit these
abundance patterns \citep[for a review see][]{Gratton+04}.
These chemical variations have been interpreted as originating from the presence of both a ``normal'' stellar population, exhibiting abundances similar to field stars of the same metallicity, and a second  population of stars formed out of material processed by hydrogen burning via the CNO-cycle and the NeNa and MgAl chains \citep[e.g.][]{Prantzos+07}.  {According to \citet{Carretta+09},  50-70\% of the stars in gloular clusters belong to the second population.}

Two sources of processed ejecta have been proposed: the slow winds
  of \emph{massive AGB stars}, which enrich their convective envelopes
  with H-burning products
  \citep{Ventura+01,Dantona+02,Denissenkov+Herwig03} and fast-rotating
  massive stars (we refer to these as \emph{spin stars}), which
  are believed to expel processed material centrifugally when they
  reach break-up rotation
\citep{Prantzos+Charbonnel06,Decressin+07a}.%
%
%
In this scenario a first generation of stars is formed out of pristine
material. Their low-velocity ejecta are trapped inside the potential
well of the cluster and provide the material for the formation of a
second generation of stars%
\footnote{\citet{Glebbeek+09} suggested that a chain of multiple
   stellar collisions in the dense center of a star cluster may also
   enrich the interstellar medium with processed material.}.

Although both proposed sources are promising, matching the observed
abundance patterns and providing enough ejecta for the formation of a
second generation which outnumbers the first generation have proven to
be two major challenges.
In this Letter we propose \emph{massive binaries} as a candidate
for the internal pollution of globular clusters.

\begin{figure}[]
 \vspace*{-2.0 cm}
\begin{center}
 \includegraphics[width=\textwidth]{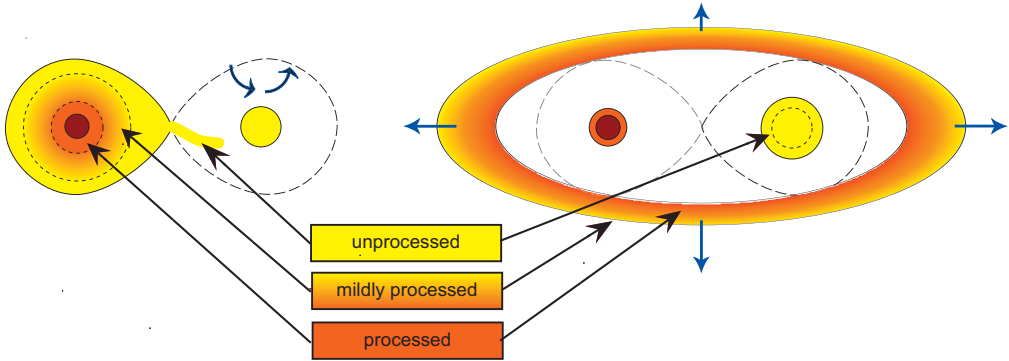} 
 \caption{Cartoon representation of the proposed scenario. {\bf Left}: A binary system at the onset of mass transfer. The deepest layers in the donor star have been processed by proton capture reactions. The accreting star spins up as it accretes mass and angular momentum until it approaches the break-up limit. {\bf Right}: The same system after the donor star has been stripped from its envelope. The companion star acrreted just a fraction of the transferred mass, mainly unprocessed material originating from the outermost layers of the donor star.  Material orginating from deeper layers in the donor star are shedded into a circumbinary disk.    }
   \label{fig1}
\end{center}
\end{figure}

\section{Binaries as sources of  enrichment}

Interacting binaries can shed large amounts of material processed by
hydrogen burning into their surroundings.  A clear example is the
well-studied system RY Scuti. It is undergoing rapid mass transfer
from a 7\Msun~supergiant to its 30\Msun~companion. Mass is lost from
the system via the outer Lagrangian points into a circumbinary disk
and a wider double toroidal nebula.  The nebula shows signatures of
CNO processing: it is enriched in helium and nitrogen and depleted in
oxygen and carbon
\citep{Smith+02,Grundstrom+07}. In contrast to the high-velocity
radiatively driven winds of massive stars, these ejecta have low
velocities. \citet{Smith+01} measure expansion velocities ranging from
30 to 70 km~s$^{-1}$ in the nebula of RY Scuti, which are smaller than
the present-day escape velocity of massive globular
clusters. Furthermore the nebula shows evidence for clumps
\citep{Smith+02} and dust in the outer parts \citep{Gehrz+01}, which
may serve as seeds for the formation of a second generation of
low-mass proto-stars.

Evidence for severe mass loss from interacting binaries comes
from a wide variety of observed interacting and post-interaction
systems: it appears to be a common phenomenon for many interacting
binaries.
Various authors have inferred highly non-conservative evolution for
Algols, systems which are currently undergoing stable mass transfer
\citep[e.g.][]{Refsdal+74,deGreve+Linnell94, Figueiredo+94,
VanRensbergen+06, DeMink+07}. Most notable are short-period
 binaries containing a compact object, e.g. cataclysmic variables,
 X-ray binaries, binary radio pulsars and double white dwarf
 systems. Their formation requires a phase of severe mass and angular
 momentum loss by ejection of a common envelope. Direct evidence for
 this type of evolution comes from planetary nebulae with close binary
 nuclei, which appear to have recently emerged from the
 common-envelope phase \citep[for a review see][]{Iben+Livio93}.

Theoretical considerations support the idea that most interacting
binaries shed large amounts of mass. Three-dimensional hydrodynamical
simulations of the mass transfer stream and accretion disk of the
interacting binary $\beta$ Lyrae predict that 50\% of the transferred
mass is lost \citep{Bisikalo+00, Nazarenko+Glazunova06A}. In addition
\citet{Ulrich+Burger76} showed that the accreting star is driven out of
thermal equilibrium and expands. This can lead to contact and strong
mass and angular momentum loss from the system
\citep{Flannery+Ulrich77}.
Furthermore, \citet{Packet81} noted that the accreting star reaches
break-up rotation after gaining only a few percent of its own mass.
Rapid rotation is found for many accreting stars in Algols
\citep{Barai+04} and this mechanism has been proposed to explain the
formation of Be-X-ray binaries \citep[e.g.][]{Pols+91}. In principle,
tides can counteract the effect of spin-up by mass transfer in close
binaries. \citet{Petrovic+05_WR} computed detailed binary evolution
models taking into account these effects. They find that massive
binaries with initial periods as short as 3-6 days lose 70-80\% of the
transferred mass on average. For wider and more massive systems they
expect even less conservative mass transfer, such that nearly the
entire envelope of the primary is returned to the interstellar medium.

\section{Composition of the ejecta\label{sec:model}} 
In \citet{DeMink+09b} we present the computations of a  the evolution of a 20\Msun~star in a close binary
     taking into account the effects of non-conservative mass and
     angular momentum transfer, rotation and tidal interaction
We find that this system sheds about
     10\Msun~of material, nearly the entire envelope of the primary
     star. The ejecta are enriched in He, N, Na and Al and depleted
     in C and O, similar to the abundance patterns observed in gobular
     cluster stars. {However, Mg is not significantly depleted in the ejecta of this model.} For a more detailed discussion we refer to \citet{DeMink+09b}.

\section{Mass budget\label{sec:pop}}

\begin{figure}[b]
\centering
\includegraphics[width=0.85\textwidth]{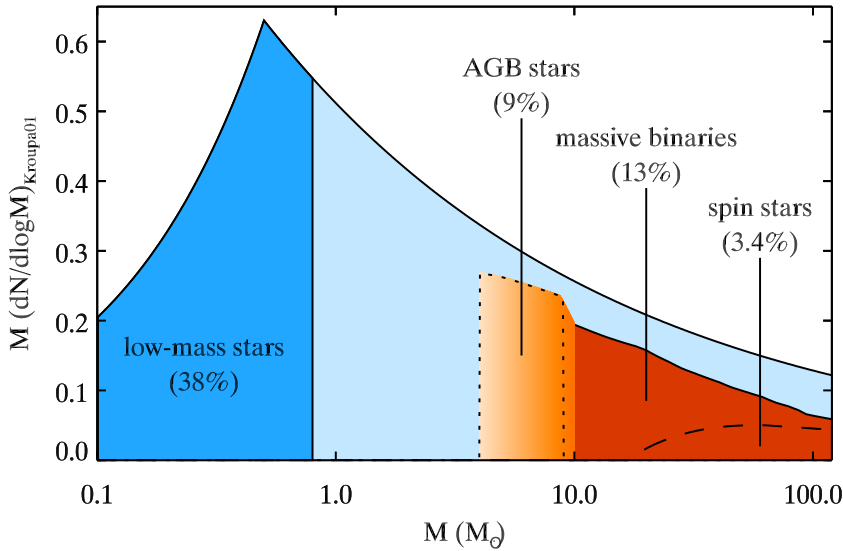}
\caption{ 
Mass weighted \citet{Kroupa01} IMF as a function of stellar mass. The
 surface areas indicate the mass contained in the first generation of
 long-lived low mass stars (dark blue), the ejecta of AGB stars
 (dotted line), spin stars, i.e. fast-rotating massive stars (dashed line) and massive (red) and
 intermediate-mass (orange) binaries. Percentages indicate the
 fraction of mass relative to the total mass contained in stars of the
 first generation. See Sec.~\ref{sec:pop} for
 details. \label{fig:pop} This Figure has been adapted from \citet{DeMink+09b}}
\end{figure}

One of the main challenges for the two previously proposed sources of
pollution, massive AGB and spin stars, is to provide the large amount
of ejecta needed to create a second population which is larger than
the first population.
The population of low-mass stars (0.1-0.8\Msun), which can still be
observed today, represents 38\% of the stellar mass initially present
in the cluster assuming a standard \citet{Kroupa01} initial mass
function (IMF) between 0.1-120\Msun, see Fig~\ref{fig:pop}.  The
ejecta of AGB stars with initial masses between 4 and 9\Msun~represent
up to  8.9\% of the initial stellar mass
\citep[assuming an initial-final mass relation by][]{Ciotti+91}.  For
spin stars this fraction is 3.4\%, if one assumes that every massive
star is single and born with a rotational velocity high enough to
reach break-up rotation \citep[using models by][]{Decressin+07a}. These ejecta are not sufficient to create a
second generation which is equally numerous as the first generation,
even when we assume that the second generation consists only of low
mass stars and that star formation is very efficient, see
Fig~\ref{fig:pop}.

Two rather extreme solutions have been proposed.
(1) The IMF was highly anomalous, favoring the formation of the
polluting stars with respect to the long-lived low-mass stars that we
observe today. Even though we have no direct constraints on the IMF of
globular clusters,
\citet{Kroupa02} finds that the IMF is remarkably uniform in stellar
populations with very different properties.
(2) Clusters were initially at least 10-20 times more massive and they
have preferentially lost low-mass stars from the first generation as a
result of the dynamical evolution and tidal stripping of the cluster
\citep{Decressin+07b, Decressin+08, D'Ercole+08}.
In this section we investigate to what extent the ejecta of massive
binaries can alleviate this conundrum.
Even though the current fraction of detected binaries in globular
clusters is not high \citep[e.g.][]{Davies+08}, this is not
necessarily the case for the high-mass stars originally present in the
cluster. \citet{Sana+08} and \citet{Mason+09} find a minimum binary
fraction of 60-75\% for O stars associated with clusters or OB
associations.  In globular clusters these fractions may even be
higher.  In this environment close binaries can be created and
tightened during and after the star formation process, for example by
the dissipative interaction with gas \citep[e.g.][]{Bonnel+Bate05} and
by three-body interactions such as the Kozai mechanism in combination
with tidal friction \citep{Fabrycky+07}.  With massive stars
preferentially residing in the dense core of the cluster, where the
dynamical encounters are most frequent, it is not unreasonable to
assume that the large majority of massive stars interact by mass
exchange.

Let us assume that every massive star is member of an interacting
binary.  In Sects.~2 we argued that nearly the entire envelope
of the donor is returned to the ISM.  For simplicity, we neglect the
contribution of the secondary star after it has been spun up by mass
transfer or during a possible phase of reverse mass transfer and we
assume that entire envelope of the primary becomes available for star
formation.
%
We assume helium core masses as in \citet{Prantzos+Charbonnel06} for
stars more massive than 10\Msun. Under these assumptions the slow
ejecta of massive binaries represent 13\% of the mass originally
present in stars: more than the ejecta of AGB and spin stars combined.


Measurements of lithium suggest that the ejecta of the first generation are diluted with pristine gas \citep{Pasquini+05}.
Together with an equal amount of pristine gas, the ejecta of binaries
with donors more massive than 10\Msun~would represent 26\% of the
initial cluster mass (compared to 38\% contained in the first
generation of low mass stars). 
The adopted lower mass limit for our binary scenario is rather arbitrary.  If we take into account the {potential} contribution of intermediate mass stars (4-10\Msun) according to this scenario, the ejecta would be sufficient to form a second population of chemically peculiar stars that outnumbers the first generation of normal stars. 
 { The assumptions in this scenario can be relaxed if the evaporation of stars from the cluster preferentially affects the first stellar generation, as suggested by \citet{Decressin+08} .  }

\section{Conclusions}
We discussed the potential of massive binaries as the source of enrichment 
in globular clusters. The majority of massive stars are
expected to be members of interacting binary systems. These return
most of the envelope of their primary star to the interstellar medium
during non-conservative mass transfer.  We show that the amount of
polluted material ejected by binaries may be larger than that of the two
previously suggested sources: massive AGB stars and the slow winds of
fast rotating massive stars. After dilution with pristine material, as
lithium observations suggest, binaries could return enough
material for the formation of a chemically enriched second generation
that is equally numerous as the first generation of low-mass stars,
without the need to assume a highly anomalous IMF, external
polution of the cluster or a significant loss of stars from the
non-enriched first generation.

In addition to providing a new source of slowly-ejected enriched
material, binary interaction also affects the previously proposed
scenarios.  Binary mass transfer naturally produces a large number of
fast-rotating massive stars which may enrich their surroundings even
further.  Binary interaction will also affect the yields of
intermediate-mass stars. Premature ejection of the envelope in
4-9\Msun~stars will result in ejecta with less pronounced
anti-correlations as suggested in the AGB scenario. On the other hand,
we expect that binary-induced mass loss may also prevent the dredge-up
of helium burning products. 

For a detailed comparison of the chemical predictions of this scenario binary models {for a range of masses and orbital periods} are needed and population synthesis models are essential to fullly evaluate the mass budget of the different sources.  {Some peculiar feature, such as the apparant presence of distinct, chemically homogeneous subpopulations in  $\omega$~Cen and NGC~2808 \citep[e.g.][]{Renzini08} deserves further attention.}

\bibliographystyle{aa}



\end{document}